\documentclass[fleqn,usenatbib]{mnras}

\usepackage{newtxtext,newtxmath}

\usepackage[T1]{fontenc}
\usepackage{ae,aecompl}

\usepackage{graphicx}	
\usepackage{amsmath}	
\usepackage{amssymb}	
\usepackage{hyperref}
\usepackage[referable]{threeparttablex}

\newcommand{\MAXI}{MAXI~J1820$+$070}
\newcommand{\MJD}[1]{MJD~$#1$}
\newcommand{\MJDI}[2]{MJD~$#1$--$#2$}
\newcommand{\ph}[1]{\phantom{#1}}

\title[Polarimetry of black hole X-ray binary \MAXI]{Disc and wind in black hole X-ray binary \MAXI\ observed through polarized  light during its 2018 outburst} 

\author[I.A. Kosenkov et al.]{Ilia A. Kosenkov,$^{1,2}$\thanks{E-mail: ilia.kosenkov@utu.fi} Alexandra Veledina,$^{1,3,4}$ Andrei V. Berdyugin,$^{1}$ Vadim Kravtsov,$^{1}$
\newauthor Vilppu Piirola,$^{1}$ Svetlana V. Berdyugina,$^{5,6}$ Takeshi Sakanoi,$^{7}$ Masato Kagitani$^{7}$
\newauthor and Juri Poutanen$^{1,3,4}$ 
\\ 
$^{1}$Department of Physics and Astronomy, FI-20014 University of Turku, Finland \\ 
$^{2}$Department of Astrophysics, St. Petersburg State University, Universitetskiy pr. 28, Peterhof, 198504 St. Petersburg, Russia\\
$^{3}$Nordita, KTH Royal Institute of Technology and Stockholm University, Roslagstullsbacken 23, SE-10691 Stockholm, Sweden\\
$^{4}$Space Research Institute of the Russian Academy of Sciences, Profsoyuznaya Str. 84/32, 117997 Moscow,  Russia\\
$^{5}$Leibniz-Institut f\"{u}r Sonnenphysik, Sch\"{o}neckstr. 6, 79104 Freiburg, Germany\\
$^{6}$Institute for Astronomy, University of Hawaii, 2680 Woodlawn Drive, Honolulu, 96822-1897 HI, USA \\
$^{7}$Graduate School of Science, Tohoku University, Aoba-ku, 980-8578 Sendai, Japan}

    \pubyear{2020}

\begin{document}
\label{firstpage}
\pagerange{\pageref{firstpage}--\pageref{lastpage}}
\maketitle
   
\begin{abstract}
We describe the first complete polarimetric dataset of the entire outburst of a low-mass black hole X-ray binary system and discuss the constraints for geometry and radiative mechanisms it imposes.
During the decaying hard state, when the optical flux is dominated by the non-thermal component, the observed polarization is consistent with the interstellar values in all filters.
During the soft state, the intrinsic polarization of the source is small, $\sim$0.15~per~cent in $B$ and $V$ filters, and is likely produced in the irradiated disc. 
A much higher polarization, reaching $\sim$0.5~per~cent in $V$ and $R$ filters, at position angle of $\sim 25\degr$ observed in the rising hard state coincides in time with the detection of winds in the system.
This angle coincides with the position angle of the jet. 
The detected optical polarization is best explained by scattering of the non-thermal (hot flow or jet base) radiation in an equatorial wind.
\end{abstract}

\section{Introduction}

The onset of accretion processes in a black hole X-ray binary (BHB) is accompanied by bright flares of emission from radio to X-ray energies.
Every year, there are a couple of new BHBs discovered.
The outburst lasts for weeks to months, and then the source typically becomes quiescent again for decades.
Radiation processes shaping the broadband emission in these objects and the source of their variability are unknown.
While there is a general consensus on the components producing emission at X-ray and radio wavelengths, the origin of optical and infrared radiation has been debated.
Proposed candidates include irradiated disc, hot accretion flow and jet \citep{Poutanen2014a,Uttley2014}.
 
\begin{figure*}
 \begin{minipage}[c]{0.67\textwidth}
    \includegraphics[width=\textwidth]{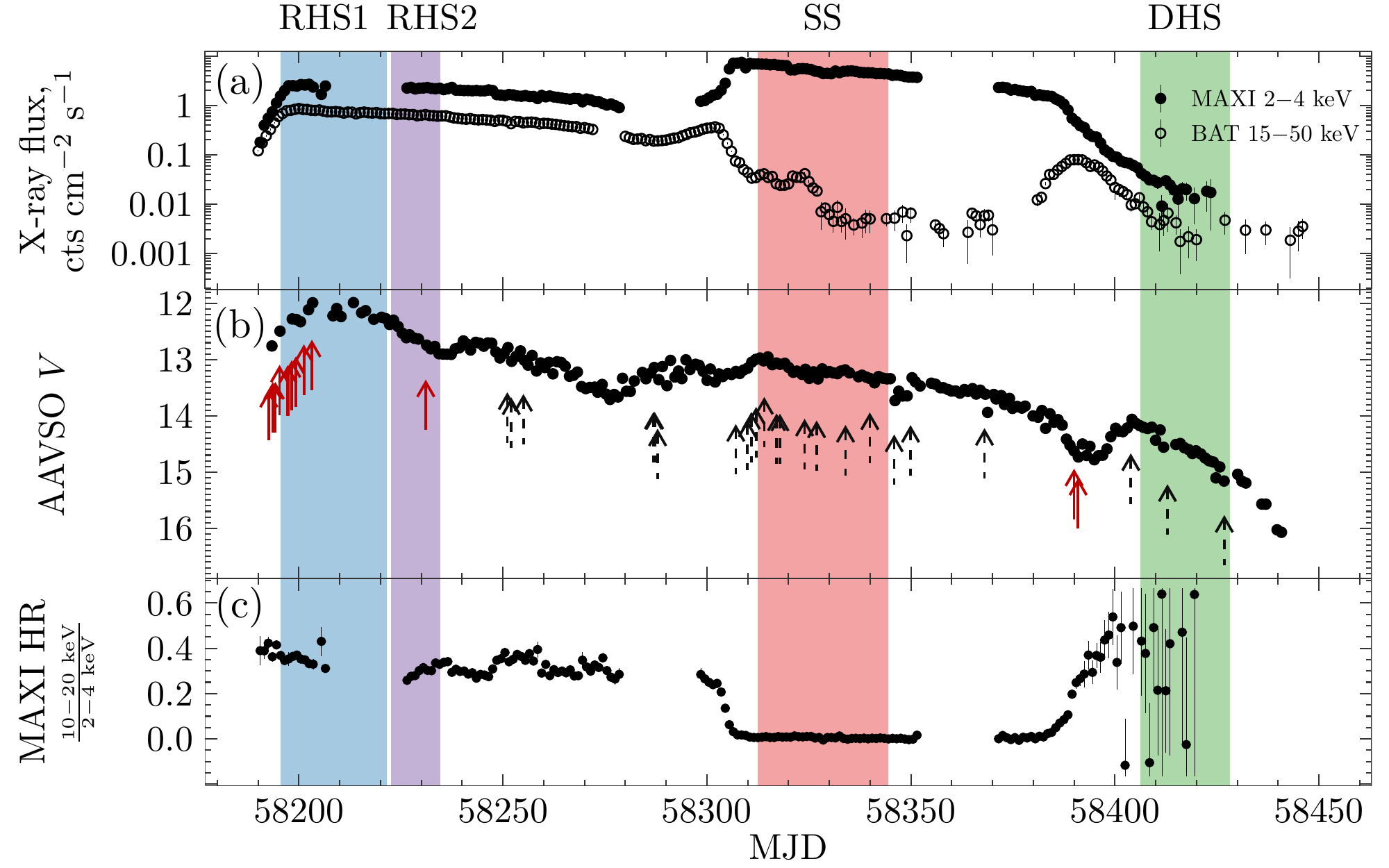}
  \end{minipage}\hfill
  \begin{minipage}[c]{0.3\textwidth}
    \caption{
    X-ray/optical light curves and hardness ratio of \MAXI. 
    (a) MAXI 2--4~keV (filled circles) and BAT 15--50~keV (open circles) fluxes, 
    (b) AAVSO Johnson $V$-filter light curves \citep[][\url{https://www.aavso.org}]{Kafka2019}, 
    (c) MAXI 10--20/2--4\,keV hardness ratio. Errors are 1$\sigma$.
    Coloured areas highlight time intervals with polarimetric observations.
    Vertical arrows indicate spectroscopic observations of \MAXI\ by \citet{MunozDarias2019}: solid red arrows denote wind detections, dashed black arrows indicate absence of wind features.
    RHS1 and 2 mark two phases of the rising hard state, SS -- the soft state, DHS -- the decaying hard state. 
    }
     \label{fig:LC}
  \end{minipage}
\end{figure*}

Several methods have been used to distinguish between these alternatives.
Studies of long-term optical and infrared (OIR) light-curves showed that there are at least two components  \citep[e.g.,][]{Russell2010,Kalemci2013,Kalemci2016,Poutanen2014,Kosenkov2020}.
Thermal component, which dominates emission in the soft state, is attributed to the irradiated disc. 
The non-thermal component, which manifests itself through bright OIR flares observed in the hard X-ray spectral state (HS), is attributed either to the hot flow or the jet.
Fast OIR variability and its relation to X-rays can be used to discriminate between these components, as the latter are expected to have different signatures in the cross-correlation function (for observations see, e.g., \citealt{Kanbach2001,Durant2008,Gandhi2010,Pahari2017}, and for modelling, e.g., \citealt{Veledina2017,Malzac2018}).
Ideally, simultaneous OIR and X-ray observations have to be performed throughout the outburst, however, such observations are difficult to acquire, and the coverage remains sparse.

Another potential possibility to study the origin of OIR emission is through the polarization signatures, which are expected to be different for different components.
In order to understand their contributions to the total spectrum at various outburst stages, high-precision polarimetric observations covering the whole outburst are needed. 
However, such observations have not been presented for any BHB so far.

The enhanced X-ray emission from \MAXI\ has been detected by the Monitor of All-sky X-ray Image \citep[{MAXI,}][]{Matsuoka2009, Mihara2011} on 2018 March 11 (\MJD{58188})  \citep{Kawamuro2018}.
The object was later identified \citep{Denisenko2018} with the optical source discovered a few days earlier by the All-Sky Automated Survey for SuperNovae \citep{ASAS-IN}.
Typical for a BHB in the outburst, the source initially demonstrated hard X-ray spectrum, which started to soften four months later, indicating the beginning of transition to the soft state (SS) \citep{Homan2018}.
On \MJD{58383}, \MAXI\ showed the reverse transition to the HS \citep{Negoro2018,Motta2018}, thus completing the standard X-ray activity cycle of a low-mass BHB.

Exceptional brightness of \MAXI\ initiated numerous observational campaigns. 
The source has been observed in OIR using fast photometric instruments \citep{Gandhi2018,Littlefield2018,Sako2018,Casella2018}.
The revealed complex interconnection of the optical and X-ray light-curves could be interpreted in terms of three components contributing to the optical emission: a hot flow giving broad dip in the cross-correlation function, a jet giving the narrow peak at subsecond optical lags, and a reprocessed disc component manifesting itself as a broad optical peak lagging X-rays \citep{Paice2019}.
These components have different predictions for polarization degree (PD) and polarization angle (PA), their spectral dependence and dependence on the accretion state.
By tracing the evolution of these signatures one can determine the source of polarized  light and its relative role in the optical emission at each stage.

In this paper, we describe the high-precision polarimetric data obtained in the SS and decaying HS (DHS) in $BVR$ filters.
Together with the reported earlier data on the rising HS (RHS) obtained with the same telescope and instrument, the described dataset represents the first study of the evolution of optical polarimetric signatures across all major X-ray spectral states.
We also present extensive polarization measurements of the field stars, which we further use to get a reliable estimate for the interstellar polarization in the source direction.

\section{Data}\label{sec:Data}

\subsection{Optical photometry and X-ray data}\label{sec:Data_XR}

We use the X-ray light-curves in 2--4 and 10--20\,keV bands obtained with MAXI\footnote{\url{http://maxi.riken.jp/pubdata/v6l/J1820+071/index.html}} to calculate the hardness ratio and to determine the spectral state of the source.
Because MAXI data have a gap during our observations in the RHS, we also use \textit{Neil Gehrels Swift Observatory} Burst Alert Telescope\footnote{\url{https://swift.gsfc.nasa.gov/results/transients/weak/MAXIJ1820p070/}} \citep[BAT,][]{Gehrels2004, Barthelmy2005} X-ray light curve in the 15--50\,keV range. 
The hard X-ray flux has been smoothly decaying during the gap.
The light-curves and evolution of the hardness ratio are shown in Fig.\,\ref{fig:LC}a,c.

The optical photometric data are obtained from the AAVSO International Database.\footnote{\url{ https://www.aavso.org}}
\MAXI\ was monitored in $V$-band throughout the 2018 outburst.
We use the Johnson $V$-band data binned to 1~day intervals (Fig.\,\ref{fig:LC}b).

\begin{table*}
\centering
\caption{The average observed and intrinsic polarization towards \MAXI\ and the estimate of the ISM polarization. Errors are 1$\sigma$.} 
\label{tbl:averages}
\begin{threeparttable}
\begin{tabular}{cr@{$\ \pm\ $}l r@{$\ \pm\ $}l r@{$\ \pm\ $}l r@{$\ \pm\ $}lr@{$\ \pm\ $}lr@{$\ \pm\ $}l}
\hline
   & \multicolumn{4}{c}{$B$} & \multicolumn{4}{c}{$V$} & \multicolumn{4}{c}{$R$} \\
 State & \multicolumn{2}{c}{$P$} & \multicolumn{2}{c}{$\theta$} & \multicolumn{2}{c}{$P$} & \multicolumn{2}{c}{$\theta$} & \multicolumn{2}{c}{$P$} & \multicolumn{2}{c}{$\theta$} \\
 & \multicolumn{2}{c}{(per cent)} & \multicolumn{2}{c}{(deg)} & \multicolumn{2}{c}{(per cent)} & \multicolumn{2}{c}{(deg)} & \multicolumn{2}{c}{(per cent)} & \multicolumn{2}{c}{(deg)} \\
\hline
\multicolumn{13}{c}{Observed polarization} \\
RHS1\tnotex{fn:a}&    0.76 &    0.01 &   53.9 &  0.3 &    0.79 &    0.01 &   54.7 &    0.4 &    0.76 &    0.01 &   53.3 &    0.3 \\
RHS2\tnotex{fn:a}&    0.76 &    0.02 &   51.4 &    0.6 &    0.87 &    0.02 &   50.5 &    0.8 &    0.86 &    0.02 &   45.8 &    0.6 \\
SS&    0.66 &    0.01 &   61.5 &    0.4 &    0.67 &    0.01 &   62.2 &    0.5 &    0.62 &    0.01 &   63.5 &    0.4 \\
DHS&    0.76 &    0.04 &   62.2 &    1.4 &    0.63 &    0.06 &   64.1 &    2.6 &    0.67 &    0.04 &   62.0 &    1.7 \\
\hline
ISM  &0.81 & 0.03 & 64.0 & 1.1 &0.71 & 0.03 & 68.4 & 1.2 &0.60 & 0.02 & 64.4 & 0.8 \\
\hline
\multicolumn{13}{c}{Intrinsic polarization} \\
RHS1 &    0.28 &    0.01 &    9.2  & \ph{1}1.0 &    0.36 &    0.01 &    22.9 & \ph{1}1.0 &    0.30 &    0.01 &   29.0 & \ph{1}0.9 \\
RHS2 &    0.34 &    0.02 &    8.8  & \ph{1}1.4 &    0.51 &    0.02 &    23.4 & \ph{1}1.4 &    0.53 &    0.02 &   23.9 & \ph{1}1.1 \\
SS   &    0.16 &    0.01 &   $-$15.8 & \ph{1}1.6 &    0.15 &    0.01 &    13.4 & \ph{1}2.3 &    0.02 &    0.01 &   39.1 &      11.0 \\
DHS  &    0.06 &    0.04 &    $-$3.0 &      15.4 &    0.13 &    0.06 &     2.8 &      12.4 &    0.09 &    0.04 &   44.8 &      12.5 \\
\hline
\end{tabular}
\begin{tablenotes}
\item[\textit{a}] \label{fn:a} Data adopted from \citet{Veledina2019}.
\end{tablenotes}
\end{threeparttable}
\end{table*}

\subsection{Optical polarimetry}

Polarimetric measurements were performed using Dipol-2 polarimeter \citep{Piirola2014} mounted on the Tohoku 60 cm telescope (T60) at Haleakala observatory, Hawaii.
Dipol-2 is a remotely operated ``double-image'' CCD polarimeter, which is capable of recording polarized images in three ($BVR$) filters simultaneously.
The innovative design of the polarimeter, where the two orthogonally polarized  images of the sky overlap on the images of the source, allows to completely eliminate the sky polarization at an instrumental stage (even if it is variable), and to achieve unprecedentedly high, up to $10^{-5}$, accuracy of target polarimetric measurements \citep{Piirola1973,Berdyugin2018_Algol,Berdyugin19,Piirola2020}.

We observed \MAXI\ for 26 nights during \MJDI{58312}{58344} and for 9 nights during \MJDI{58406}{58428}.
Each night we obtained from 12 to 46 individual measurements of Stokes $q$ and $u$ parameters simultaneously in three filters $BVR$. 
To calibrate our instrument, we observed 30 unpolarized  standard stars.
The zero point of PA was determined by observing high-polarized  standards HD204827, HD25443, and HD236928.
The determined instrumental PD at the telescope, $\leq 0.005$~per~cent, is negligible in the present case.

The individual measurements were used to compute nightly average values using the ``2$\sigma$-weighting algorithm''.
The algorithm iteratively filters out outliers, assigning smaller weights to these measurements.
The errors on the Stokes $q$ and $u$ parameters were computed as standard errors of the weighted means. 
These errors were then used to estimate errors on the PD $P$ and PA $\theta$ \citep[see further details on the reduction and calibration procedure in][]{Kosenkov2017, Piirola2020}.
The nightly average PD and PA are presented in the online Supplementary material.
Based on the X-ray spectral state of \MAXI\ at the time of observations, we split all individual measurements into two groups, corresponding to the SS and DHS.
Average SS and DHS polarization data were obtained by applying the same weighting algorithm to all measurements within the group and are given in Table\,\ref{tbl:averages}.
Because \MAXI\ was brighter during the SS as compared to the DHS \citep[see, e.g.,][ and light curves in Fig.\,\ref{fig:LC}]{Russell2019}, the errors are systematically smaller.
The polarization measurements from \citet{Veledina2019}, obtained during the RHS and divided into two groups (RHS1 and RHS2), are also given for completeness.

\subsection{Interstellar polarization estimate}\label{sec:Data_IP}

The interstellar medium (ISM) located between us and the source may polarize its intrinsic emission, and we need to subtract ISM contribution from the observed polarization.
The ISM polarization can be estimated using the polarization of field stars surrounding the source.
This method has already been used to measure intrinsic polarization of \MAXI\ in the RHS \citep{Veledina2019}.
Because the polarization of the field stars does not vary, we use their polarimetric data obtained throughout the outburst of \MAXI.
We additionally obtained over 400 new measurements of the Stokes $q$ and $u$ parameters for the field star that has the smallest angular separation from  \MAXI\ and is located at a similar distance \citep[star \#2, see table~2 and figure~7 in][]{Veledina2019}.
We find that its polarization is within the error circle from other stars with similar distance and angular separation, and hence we take five field stars \citep[\#2, 3, 6, 7, and 9 from][]{Veledina2019} to obtain a reliable estimate of the ISM polarization towards \MAXI.
We calculate the weighted average of their Stokes parameters and call it the ISM polarization hereafter (Table~\ref{tbl:averages}).

\section{Results}\label{sec:Results}

The PD of \MAXI\ measured during the RHS is similar to that of the ISM, however, the PA is different.
The observed polarization increased around \MJD{58222} when the optical fluxes significantly dropped \citep{Townsend2018}.
The most prominent increase occurred in $R$ filter, where the drop of the non-thermal component is largest. 

After the transition to the SS, the PD decreased and the PA increased, becoming closer to the ISM values (see Table~\ref{tbl:averages}).
The source exhibited $\sim$0.09~mag variability in $V$ filter with a period a few per cent longer than the orbital period \citep[the latter has been measured in quiescence,][]{Torres2019}, likely originating from superhumps.
Using Lomb-Scargle power density estimate \citep{Scargle1982}, we checked whether the PD or PA show any periodicities, but did not find any significant peaks above the noise level.

\begin{figure}
\centering 
    \includegraphics[keepaspectratio, width = 0.8\linewidth]{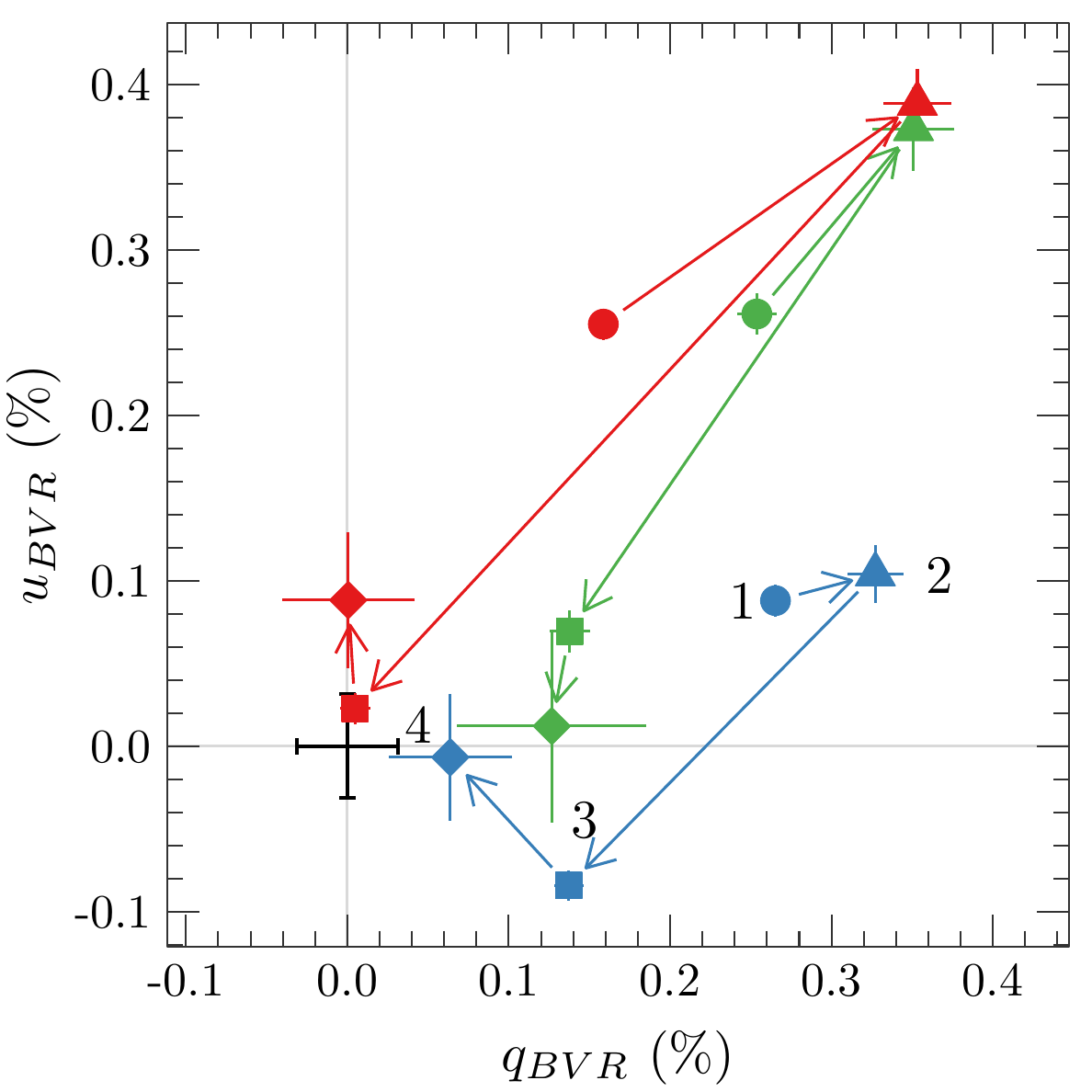}
    \caption{
        Average intrinsic Stokes $q$ and $u$ parameters of \MAXI. 
        Circles, triangles, squares and diamond correspond to the RHS1 (labelled 1),  RHS2 (2), SS (3), and DHS (4), respectively.
        Blue, green and red colour encodes $B$, $V$ and $R$ filter polarization, respectively.
        Black cross at the origin shows the typical errors in the ISM polarization.
        Error bars denote 1$\sigma$ statistical errors.
        }
    \label{fig:QU}
\end{figure}

We computed the intrinsic polarization of the source by subtracting the ISM Stokes parameters from the observed ones (Table~\ref{tbl:averages}).
In Fig.\,\ref{fig:QU} we show the intrinsic Stokes $q$ and $u$ parameters for different outburst stages.
The highest intrinsic PD has been detected during the RHS2.
The PD decreased as the source entered the SS, approaching the ISM value in $R$ filter (origin of coordinates in Fig.\,\ref{fig:QU}).
We note that the vectors of changes between RHS2 and the SS on the $(q,u)$ plane are parallel (within errors).
Finally, after the reverse transition to the DHS, we observe a small change in polarization, which is most notable in the $B$ filter. 

It is evident that there is a statistically significant difference between the RHS2 and SS Stokes parameters in all filters (see Table\,\ref{tbl:averages} and Fig.\,\ref{fig:QU}).
For other states, to judge on the significance of the detection and evolution of polarization signatures, we used the \citet{Hotelling1931} $T^2$-test (which extends Student's $t$-test onto multivariate case, see \citealt{Kosenkov2017}).
The probabilities $p$ that the Stokes parameters of the RHS1 and 2 are the same are below $2.0 \times 10^{-3}$ \citep{Veledina2019}.
The difference between the SS and DHS is not statistically significant in $V$ and $R$ filters (probabilities $p$ that they are equal are 0.05 and 0.3, respectively), but the difference is significant in $B$-filter ($p = 7 \times 10^{-5}$). 
A non-zero intrinsic polarization in the SS is highly significant in the $B$ and $V$ filters (probabilities for zero polarization are $1 \times 10^{-27}$ and $4 \times 10^{-6}$, respectively), while in $R$ filter the PD is consistent with zero with probability of 0.24.
The PD in the DHS is consistent with zero in all filters with the $2\sigma$ upper limit of about 0.1 per cent.

\section{Discussion and summary}
\label{sec:Discussion}

In this paper, we described new polarimetric observations of the SS and DHS of the black hole binary \MAXI.
Together with the reported earlier data on the RHS, this dataset presents the first detailed study of the evolution of optical polarization during the entire outburst of a BHB.
We performed extensive study of polarization of the field stars around \MAXI\ and obtained a reliable estimate for the ISM polarization towards the source (Table~\ref{tbl:averages}).
We found a small, less than 1~per~cent, intrinsic polarization of the source in the RHS and SS (see Fig.\,\ref{fig:QU}). 
On the other hand, polarization during the DHS is consistent with the ISM values.

In the SS, when the non-thermal component is quenched and the winds were not detected (see \citealt{MunozDarias2019} and Fig.\,\ref{fig:LC}), the observed emission and polarization are expected to be dominated by the irradiated disc \citep[e.g.][]{Poutanen2014,Kosenkov2020}. 
Thus the detected in this state intrinsic PD of $\sim$0.15 per cent in $B$ and $V$ and the consistent with zero PD in $R$ implies dominance of absorption in the red part of the spectrum and an increasing role of electron scattering in the blue part. 
A peculiar spectral dependence of the PA, which changes from about $-$15\degr\ in $B$ to 15\degr\ in $V$ filter, can result from a warped or flared disc, where fluxes at different wavelengths are produced by the disc annuli with different orientations with respect to the observer. 
 
In the DHS, the intrinsic PD is consistent with zero. 
The optical emission of BHB in the HS may consist of three components: thermal radiation from the irradiated disc and the non-thermal power-law-like emission from the hot flow and the jet \citep{Russell2013_J1836,Poutanen2014,Kosenkov2020}. 
Because the non-thermal components dominate in this state and the DHS polarization is negligible, we conclude that these components are intrinsically unpolarized.
Zero polarization of the non-thermal component implies tangled magnetic field or strong Faraday depolarization, both consistent with the hot flow scenario, but the base of the jet is also feasible.

We find that the high intrinsic polarization observed at the RHS (Fig.\,\ref{fig:QU}) coincides with the periods of spectroscopic detection of winds from \MAXI\ \citep[see][and Fig.\,\ref{fig:LC}]{MunozDarias2019}.
On the other hand, after the reverse transition to the DHS, the wind features were observed until \MJD{58390}, and were not detected on/after \MJD{58403}, three days before the beginning of our polarimetric run. 
The observed correlation between high PD and presence of the winds thus strongly indicates that polarization is produced by scattering of the central source radiation in the wind.
The PD of irradiated disc is rather small, as seen in the SS data, further indicating that the disc cannot be responsible for the high PD observed in the RHS.
The strong dependence of the PD on the wavelength in the RHS2, with intrinsic PD in $R$ being twice as large as in $B$ filter (see Fig.\,\ref{fig:QU}), constrains the source of the incident non-thermal radiation being the hot flow or base of the jet, because a smaller PD of the irradiated disc dilutes polarization more in the $B$ filter.

Geometry of the scattering medium can be constrained by the PA of the RHS polarization.
As the $R$-filter polarization is least diluted by the irradiated disc, we take its PA=$24\degr \pm 1\degr$ (see also fig.~12 in \citealt{Veledina2019}).
Interestingly, this is consistent with the position angle of the radio jet $25\fdg8 \pm 4\fdg4$ (computed from the coordinates of ejections, see \citealt{Bright2020}) and the X-ray jet position angle of $25\fdg1 \pm 1\fdg4$ \citep{Espinasse2020}.  
Such polarization can be produced in the following scenarios: (1) intrinsic jet synchrotron emission in the magnetic field perpendicular to the jet axis; (2) scattering of the central compact source radiation by relativistic, between 10 and 70 per cent of the speed of light, outflow \citep{Beloborodov1998} or (3) by a ring of matter in equatorial plane.

To comply with the zero DHS polarization, the first possibility requires dramatic change of the magnetic field geometry between the RHS and DHS.
Similar argument can be used against the second possibility, as it requires substantial decrease of the scattering optical depth from the RHS to DHS.
Association of a high polarization with the winds is naturally related to the third alternative.
Here, the non-thermal, hot flow or base of the jet, emission is scattered by the equatorial wind, and the polarization is detected whenever the winds are observed.
The PD of the scattered radiation in this geometry can be computed analytically \citep[see sect. 5.3 in][]{ST85} and depends only on the system inclination $i$:
\begin{equation}
    P = \frac{1-\cos^2 i}{3-\cos^2 i}.
\end{equation}
For $i=75\degr$ \citep{Torres2019}, less than two per cent of the central source radiation should be scattered in the wind to get the observed PD.

To summarise, we discovered polarization variability over the course of the outburst of BHB  \MAXI. 
The small (0.2 per cent) PD in the SS can be associated with the irradiated disc. 
We demonstrated that the non-thermal component dominating OIR flux in the HS is intrinsically unpolarized  (as observed in the DHS).
A higher, 0.3--0.5 per cent, polarization detected in the RHS, with the PA coinciding with position angle of the jet, likely originates from scattering of the non-thermal central compact source (hot flow or jet base) radiation in an equatorial wind.

\section*{Acknowledgements}

We acknowledge support from the Magnus Ehrnrooth foundation, the Academy of Finland grants 309308 and 321722 (AV), the Russian Science Foundation grant 20-12-00364  (AV, VK, JP), and ERC Advanced Grant HotMol ERC-2011-AdG-291659 (SVB, AVB).
The Dipol-2 polarimeter was built in cooperation by the University of Turku, Finland, and the Leibniz Institut fuer Sonnenphysik, Germany, with support from the Leibniz Association grant SAW-2011-KIS-7. 
We are grateful to the Institute for Astronomy, University of Hawaii, for the allocated observing time. 
We acknowledge with thanks the variable star observations from the AAVSO International Database contributed by observers worldwide and used in this research.

\bibliographystyle{mnras}
\bibliography{references}

\label{lastpage}
\end{document}